\documentclass[a4paper,11pt]{article}

\usepackage{graphicx}
\usepackage{bm}

\usepackage[utf8]{inputenc}
\usepackage[T1]{fontenc}

\usepackage[british]{babel}

\usepackage{authblk}

\usepackage{amsmath,amssymb,mathtools}
\usepackage{hyperref}
\usepackage[numbers]{natbib}

\usepackage{aas_macros}

\newcommand{\conj}[1]{\overline{#1}}
\newcommand{\R}{\bm{\mathrm{R}}}
\newcommand{\D}{\mathrm{d}}
\newcommand{\E}{\mathrm{e}}
\newcommand{\I}{\mathrm{i}}

\begin{document}


\title{The Spectral Representation of Homogeneous
Spin-Weighted Random Fields on the Sphere}

\author{%
Nicolas Tessore\thanks{%
Email: \nolinkurl{nicolas.tessore@manchester.ac.uk}}}

\affil{%
Jodrell Bank Centre for Astrophysics,
University of Manchester,\\
Alan Turing Building,
Oxford Road,
Manchester, M13 9PL, UK}

\maketitle

\begin{abstract}\noindent
This is a direct computation of the spectral representation of homogeneous spin-weighted spherical random fields with arbitrary integer spin.
It generalises known results from Cosmology for the spin-2 Cosmic Microwave Background polarisation and Cosmic Shear fields, without decomposition into $E$- and $B$-modes.
The derivation uses an instructive representation of spin-weighted spherical functions over the Spin(3) group, where the transformation behaviour of spin-weighted fields can be treated more naturally than over the sphere, and where the group nature of Spin(3) greatly simplifies calculations for homogeneous spherical fields.
It is shown that i) different modes of spin-weighted spherical random fields are generally uncorrelated, ii) the usual definition of the power spectrum generalises, iii) there is a simple relation to recover the correlation function from the power spectrum, and iv) the spectral representation is a sufficient condition for homogeneity of the fields.
\end{abstract}

\section{Introduction}

Consider a pair of (not necessarily distinct) homogeneous random fields~$X$ and~$X'$ with respective integer spins~$s$ and~$s'$ defined over the sphere.
Such fields are regularly encountered in Cosmology, where they appear as the \emph{Cosmic Microwave Background} polarisation field (spin~2) and temperature field (spin~0), as well as the \emph{Cosmic Shear} field (spin~2).
The importance of these homogeneous spin-weighted spherical random fields (SWSRF) is owed to the fact that their second-order statistics are powerful probes of the history of the universe, as their auto- and cross-correlations can be measured from observations and compared to theoretical predictions \citep{2006ewg3.rept.....P}.

In order to make such predictions, it is usually necessary to work with the \emph{spectral representation} of these random fields:
That is an expansion of the SWSRF into spin-weighted spherical harmonics (SWSH), similar to the usual spherical harmonic expansion of a scalar-valued function \cite{1966JMP.....7..863N},
\[
	X(\theta, \phi)
	= \sum_{l = 0}^{\infty} \sum_{m = -l}^{l} X_{lm} \, {}_sY_{lm}(\theta, \phi) \;,
\]
where ${}_sY_{lm}$ denotes the SWSH, and the modes~$X_{lm}$ of the random field are themselves random variables.
For any combination of $X$ and $X'$ among the temperature, polarisation, or shear fields, it has been shown that the modes~$X_{lm}$ and $X'_{l'm'}$ are necessarily uncorrelated if their numbers~$l$ or~$m$ differ,
\[
	\bigl\langle X_{lm} \conj{X'_{l'm'}} \bigr\rangle
	= \delta_{ll'} \, \delta_{mm'} \, C_l \;,
\]
where~$C_l$ is the \emph{power spectrum}, the overbar denotes complex conjugation, and the expectation~$\langle \,\cdot\, \rangle$ is taken over realisations of the random fields.
Formally, this was achieved by a decomposition of the spin-2 fields into scalar components such as $E$- and $B$-modes \citep{1997PhRvL..78.2054S,1997PhRvD..55.7368K,2002ApJ...568...20C,2002A&A...389..729S}, and using the spectral representation of homogeneous scalar-valued random fields on the sphere~\citep{yaglom1961}.

Here, it is shown by direct calculation that any pair of homogeneous SWSRF with arbitrary integer spins has a spectral representation of the above form.
This is not a trivial generalisation of the spin-0 case, because spin-weighted spherical functions (SWSF) with nonzero spin are not fully defined over the sphere alone, but additionally require in each point a reference frame in which the spin is measured.
This complication is well-known in Cosmology, where the two-point statistics of SWSRF are explicitly made frame-independent by local rotations of the spin fields, which requires careful geometric considerations \citep{1997PhRvD..56..596H}.
However, \citet{2016JMP....57i2504B} has recently pointed out that a natural domain for SWSF and SWSH exists in the Spin(3) group, the space of three-dimensional rotations.
There, not only is the additional structure of the fields due to their spin-weight intrinsically taken into account:
It also turns out that the group structure of Spin(3) greatly simplifies calculations when dealing with homogeneous random fields.

The outline of this note is as follows.
Section~\ref{sec:swsh} introduces SWSF and SWSH over Spin(3) as a natural extension of the definition over the sphere.
Section~\ref{sec:swsrf} defines the correlation function of homogeneous SWSRF over Spin(3) and relates it to the usual correlation function.
The short derivation of the spectral representation, as well as a number of related results, are given in Section~\ref{sec:sr}.
Finally, the main results are discussed in Section~\ref{sec:dis}, and two related results for practical applications are given in the appendix.

\section{Spin-Weighted Spherical Harmonics}
\label{sec:swsh}

\begin{figure}[t!]%
\centering%
\includegraphics{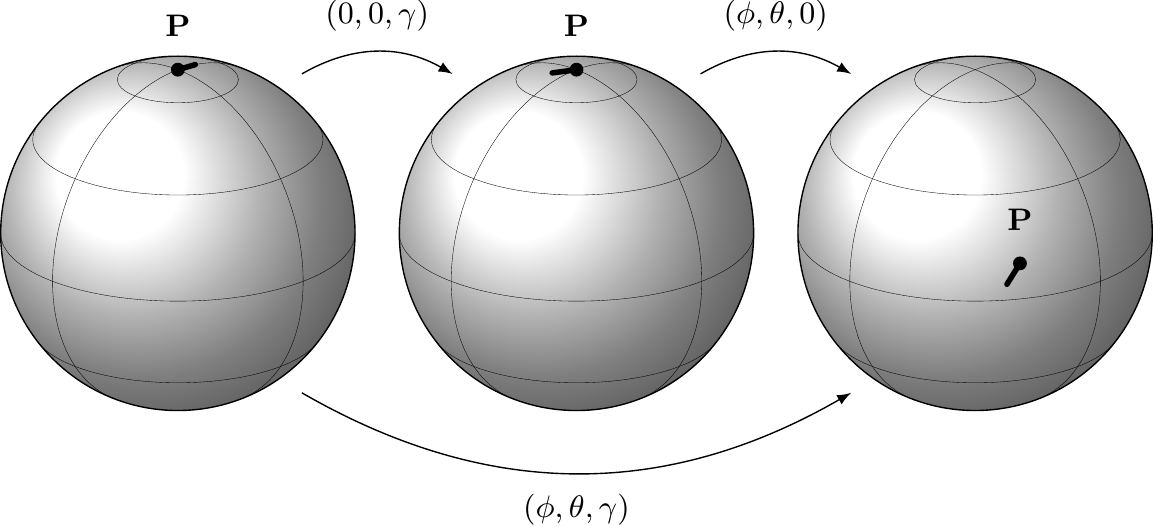}%
\caption{%
Effect of the rotation~$(\phi, \theta, \gamma)$ on the local coordinate frame in~$P$.
}%
\label{fig:rot1}%
\end{figure}

Because the value of a spin-weighted function in a given point is generally not invariant under rotations, SWSF are not fully defined on the sphere alone, but depend on a further angle~$\gamma$ that describes a reference direction in each point against which the spin is measured \citep{2016JMP....57i2504B}.
A more natural domain for SWSF is the Spin(3) group, the elements of which can be understood as the rotations~$\R = \R(\phi, \theta, \gamma)$ in three-dimensional space, written here using Euler angles~$(\phi, \theta, \gamma)$ in $z$-$y$-$z$ convention.
Over Spin(3), the transformation behaviour of a spin-weighted function is fully described by the function definition, in the following way.

On the sphere, the rotation~$\R(\phi, \theta, \gamma)$ first rotates the local coordinate frame at the pole by~$\gamma$, and then moves the rotated frame to~$(\theta, \phi)$.
This is equivalent to rotating the local coordinate frame in~$(\theta, \phi)$ by $\gamma$ (Fig.~\ref{fig:rot1}).
To extend a function~$f(\theta, \phi)$ with spin weight~$s$ from the sphere to Spin(3), it suffices to take such a rotation of the coordinate frame in each point into account,
\begin{equation}\label{eq:fn-rel}
	f(\R)
	= \E^{-\I \, s \, \gamma} \, f(\theta, \phi) \;,
\end{equation}
which is simply the definition of the spin weight of a function through its transformation behaviour under rotations \citep{2013PhRvD..87j4006B}.
With this relation, SWSF are naturally extended from the sphere to Spin(3), and the dependency on the local orientation of the reference frame is made explicit.

Similar to the spherical harmonic expansion of scalar-valued functions over the sphere, there is an expansion of SWSF into SWSH with given spin weight~$s$ \cite{1966JMP.....7..863N},
\begin{equation}\label{eq:ex-s2}
	f(\theta, \phi)
	= \sum_{l,m} f_{lm} \, {}_sY_{lm}(\theta, \phi) \;,
\end{equation}
where the modes~$f_{lm}$ of the expansion are found by integration against the complete set of SWSH,
\begin{equation}\label{eq:mo-s2}
	f_{lm}
	= \int \! f(\Omega) \, \conj{{}_sY_{lm}(\Omega)} \, \D\Omega \;,
\end{equation}
and the integral is over the sphere with the usual area element~$\D\Omega$.
After multiplying both sides of the expansion~\eqref{eq:ex-s2} by~$\E^{-\I \, s \, \gamma}$, the expression on the left-hand side recovers the SWSF relation~\eqref{eq:fn-rel}, and the right-hand side is the same relation for the special case of the SWSH,
\begin{equation}\label{eq:swsh-rel}
	{}_sY_{lm}(\R)
	= \E^{-\I \, s \, \gamma} \, {}_sY_{lm}(\theta, \phi) \;,
\end{equation}
so that an equivalent SWSH expansion can be written for SWSF extended to Spin(3),
\begin{equation}\label{eq:ex}
	f(\R)
	= \sum_{l,m} X_{lm} \, {}_sY_{lm}(\R) \;.
\end{equation}
The modes of the SWSH expansion over Spin(3) are found as usual by integrating against the basis functions,
\begin{equation}\label{eq:mo}
	f_{lm}
	= \int \! f(\R) \, \conj{{}_sY_{lm}(\R)} \, \D\R \;,
\end{equation}
where the integral is defined in terms of the Haar measure of the Spin(3) group,\footnote{%
The normalisation was chosen so that the measure reduces to the usual area element of the sphere for integrands that are independent of~$\gamma$.}
\begin{equation}\label{eq:hm}
	\int \! \D\R
	= \frac{1}{4 \pi} \int_{0}^{2\pi} \! \D\phi \int_{0}^{\pi} \! \D\theta \sin\theta \int_{0}^{4\pi} \! \D\gamma \;.
\end{equation}
Importantly, the modes~$f_{lm}$ of the expansion over Spin(3) are the same as those over the sphere.
This is seen by inserting the SWSF relation~\eqref{eq:fn-rel} and SWSH relation~\eqref{eq:swsh-rel} into the definition~\eqref{eq:mo} of the Spin(3) modes, where the previously introduced factors of $\E^{-\I \, s \, \gamma}$ cancel, and carrying out the trivial integration over~$\gamma$ using the Haar measure~\eqref{eq:hm}.
The remaining integral recovers the original definition~\eqref{eq:mo-s2} of the modes on the sphere.

There are a number of properties that follow from the definition~\eqref{eq:swsh-rel} of the SWSH over Spin(3).
For example, rewriting the SWSH over the sphere in terms of the Wigner $D$-matrices \citep{1967JMP.....8.2155G}, which are intrinsically defined over the space of rotations, the SWSH over Spin(3) become a simple rescaling of the Wigner $D$-matrices \citep{2016JMP....57i2504B},
\begin{equation}\label{eq:swsh-def}
	{}_sY_{lm}(\R)
	= (-1)^s \, \sqrt{\frac{2l + 1}{4\pi}} \, D^l_{m,-s}(\R) \;.
\end{equation}
By this relation, the SWSH naturally inherit the transformation law of the Wigner $D$-matrices under compositions of rotations~$\R$ and~$\R'$ \citep{2016JMP....57i2504B,Biedenharn:1984ex},
\begin{equation}\label{eq:swsh-law}
	{}_sY_{lm}(\R \R')
	= \sum_{m'} D^l_{mm'}(\R) \, {}_sY_{lm'}(\R') \;.
\end{equation}
Similarly, the orthogonality of the SWSH over Spin(3) follows immediately from the orthogonality of the Wigner $D$-matrices,
\begin{equation}
	\int \! {}_sY_{lm}(\R) \, \conj{{}_{s'}Y_{l'm'}(\R)} \, \D\R \\
	= \delta_{ss'} \, \delta_{ll'} \, \delta_{mm'} \;.
\end{equation}
These identities are very useful in carrying out calculations involving the SWSH expansion, such as below.

However, the important result here is that the modes~\eqref{eq:mo-s2} of the SWSH expansion over the sphere are the same as the modes~\eqref{eq:mo} of the expansion over Spin(3), so that the latter representation can be used to obtain results about the former.
This is enormously helpful in performing calculations in the context of homogeneous random fields, since their symmetry operations (rotations of the sphere) coincide with the translations of the Spin(3) group.

\section{Spin-Weighted Spherical Random Fields}
\label{sec:swsrf}

\begin{figure}[t!]%
\centering%
\includegraphics{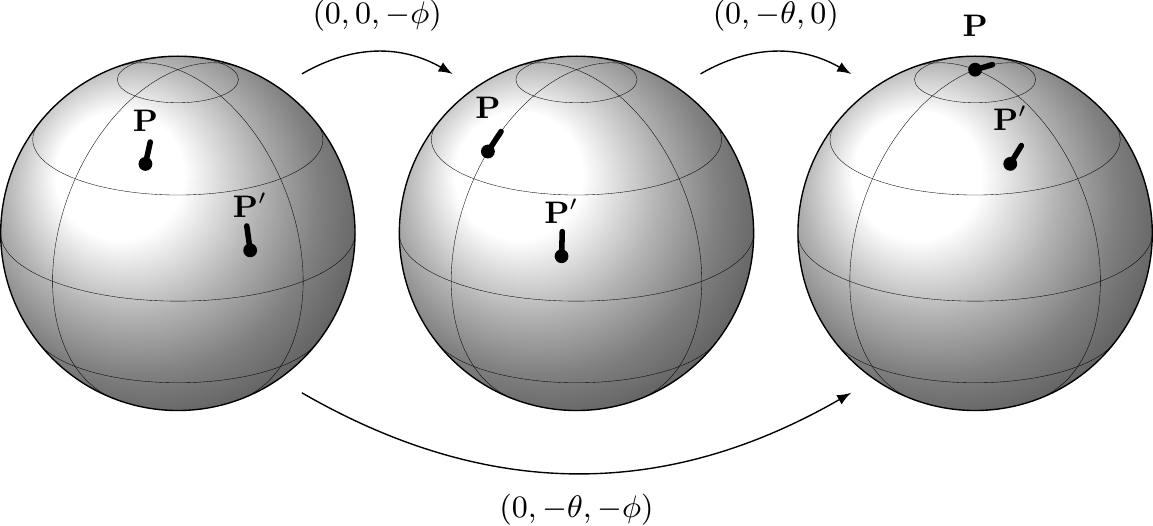}%
\caption{%
Rotations of the points $P = (\theta, \phi)$ and $P' = (\theta', \phi')$ and their local coordinate frames on the sphere.
}%
\label{fig:rot2}%
\end{figure}

The SWSF are now replaced by homogeneous random fields~$X$ and~$X'$ as initially discussed.
For SWSRF defined over Spin(3), homogeneity requires the covariance between $X(\R)$ and $X'(\R')$ to be invariant under arbitrary rotations~$\R_0$,\footnote{%
The random fields~$X$ and~$X'$ in the definition~\eqref{eq:hom} of homogeneity must be ``jointly homogeneous'' to be invariant under simultaneous rotations of both fields.
This is not always the case:
For example, if $X'$ is obtained by rotating $X$ a certain amount about an arbitrary axis, both $X$ and $X'$ are individually homogeneous, but not jointly so.}
\begin{equation}\label{eq:hom}
	\bigl\langle X(\R) \, \conj{X'(\R')} \bigr\rangle
	= \bigl\langle X(\R_0 \R) \, \conj{X'(\R_0 \R')} \bigr\rangle \;.
\end{equation}
In particular, the invariance must also hold for $\R_0 = \R^{-1}$, in which case the right-hand side only depends on the product $\R^{-1} \R'$, but neither $\R$ nor $\R'$ individually.
This suggests the definition of the correlation function~$\xi$ over Spin(3),
\begin{equation}\label{eq:xi-spin3}
	\bigl\langle X(\R) \, \conj{X'(\R')} \bigr\rangle
	\equiv \xi(\R^{-1} \R') \;.
\end{equation}
However, in order to make the definition workable, it will also be necessary to find a simple functional form for $\xi(\R)$.

Through the SWSF relation~\eqref{eq:fn-rel}, the definition~\eqref{eq:xi-spin3} can be equivalently expressed in terms of the fields over the sphere,
\begin{equation}\label{eq:xi-spin3-to-s2}
	\xi(\R^{-1} \R')
	= \E^{-\I \, s \, \gamma} \, \bigl\langle X(\theta, \phi) \, \conj{X'(\theta', \phi')} \bigr\rangle \, \E^{\I \, s' \, \gamma'} \;.
\end{equation}
To understand how the argument on the left-hand side relates to the angles on the right, it can be expanded into a series of elementary rotations,
\begin{equation}
\begin{split}
	\R^{-1} \R'
	&= \R(-\gamma, -\theta, -\phi) \, \R(\phi', \theta', \gamma') \\
	&= \R(-\gamma, 0, 0) \, \R(0, -\theta, -\phi) \, \R(\phi', \theta', 0) \, \R(0, 0, \gamma') \\
\end{split}
\end{equation}
The factor $\R(0, -\theta, -\phi) \, \R(\phi', \theta', 0)$ describes the relative orientation of the points and their local reference frames on the sphere (Fig.~\ref{fig:rot2}), and can be written as a single rotation~$\R(\alpha, \beta, -\alpha')$ to simplify the expression,
\begin{equation}
\begin{split}
	\R^{-1} \R'
	&= \R(-\gamma, 0, 0) \, \R(\alpha, \beta, -\alpha') \, \R(0, 0, \gamma') \\
	&= \R(\alpha - \gamma, \beta, \gamma' - \alpha') \;.
\end{split}
\end{equation}
The relation~\eqref{eq:xi-spin3-to-s2} can therefore be written with explicit arguments using the newly introduced angles,
\begin{equation}
	\xi(\alpha - \gamma, \beta, \gamma' - \alpha')
	= \E^{-\I \, s \, \gamma} \, \bigl\langle X(\theta, \phi) \, \conj{X'(\theta', \phi')} \bigr\rangle \, \E^{\I \, s' \, \gamma'} \;,
\end{equation}
so that the left-hand side is brought into the desired form of a simple function definition by the substitutions $\gamma \mapsto \alpha - \gamma$ and $\gamma' \mapsto \alpha' + \gamma'$,
\begin{equation}\label{eq:xi-func-rel}
	\xi(\gamma, \beta, \gamma')
	= \E^{\I \, s \, (\gamma - \alpha)} \, \bigl\langle X(\theta, \phi) \, \conj{X'(\theta', \phi')} \bigr\rangle \, \E^{\I \, s' \, (\alpha' + \gamma')} \;.
\end{equation}
The right-hand side of the resulting expression contains the definition of the correlation function for homogeneous SWSRF over the sphere,
\begin{equation}\label{eq:xi-s2}
	\E^{-\I \, s \, \alpha} \, \bigl\langle X(\theta, \phi) \, \conj{X'(\theta', \phi')} \bigr\rangle \, \E^{\I \, s' \, \alpha'}
	\equiv \xi(\beta) \;,
\end{equation}
where the angles introduced above are now understood to be the angular distance~$\beta$ between the points (i.e.\@ the central angle), and the angles $\alpha, \alpha'$ that rotate the local coordinate frames of $X, X'$ to align with the geodesic (i.e.\@ the great circle) connecting the points \citep{2004MNRAS.350..914C,1999IJMPD...8...61N,1997PhRvD..56..596H}.\footnote{%
In weak gravitational lensing, the rotation of the spin-weighted field is usually implicit in the decomposition of the shear into tangential and cross-components.}
Explicit expressions for the angles are given in Appendix~\ref{sec:ang}.

Substituting the correlation function~\eqref{eq:xi-s2} in the expression~\eqref{eq:xi-func-rel}, and writing the result in terms of the angles of a rotation~$\R = \R(\phi, \theta, \gamma)$, yields the general expression for the correlation function over Spin(3),
\begin{equation}\label{eq:xi-rel}
	\xi(\R)
	= \xi(\phi, \theta, \gamma)
	= \E^{\I \, s \, \phi} \, \xi(\theta) \, \E^{\I \, s' \, \gamma} \;.
\end{equation}
The correlation function over Spin(3) is therefore not real-valued, but has both a spin weight of~$s'$ and an ``azimuthal spin weight'' of~$s$, while the actual amount of correlation is described by the usual real-valued function~$\xi(\theta)$ of angular distance.
Interestingly, this is precisely of the same form as the relation between the Wigner (large) $D$-matrix and Wigner (small) $d$-matrix given below.

\section{The Spectral Representation}
\label{sec:sr}

The spectral representation of homogeneous SWSRF can now be found with relative ease using the structure of the Spin(3) group \citep{Yaglom:1987wt}.
The modes~$X_{lm}$ of a zero-mean random field are zero-mean random variables; this is seen by taking the expectation of their definition~\eqref{eq:mo}.
The covariance between the modes~$X_{lm}$ and~$X'_{l'm'}$ is therefore $\langle X_{lm} \, \conj{X'_{l'm'}} \rangle$, and after inserting their definition~\eqref{eq:mo}, the expectation over realisations of the random fields can be moved into the integral, which yields the definition~\eqref{eq:xi-spin3} of the correlation function over Spin(3),
\begin{equation}
	\bigl\langle X_{lm} \, \conj{X'_{l'm'}} \bigr\rangle
	= \iint \! \xi(\R') \, \conj{{}_sY_{lm}(\R)} \, {}_{s'}Y_{l'm'}(\R \R') \, \D\R \, \D\R' \;,
\end{equation}
after a further substitution $\R' \to \R \R'$ via the rotation invariance of the Haar measure.
With the product of rotations contained in the second SWSH, the transformation law~\eqref{eq:swsh-law} factorises the integrals in~$\R$ and~$\R'$, and rewriting the SWSH in terms of the Wigner $D$-matrix~\eqref{eq:swsh-def} makes it possible to carry out the integration in $\R$ using orthogonality,
\begin{equation}
	\bigl\langle X_{lm} \, \conj{X'_{l'm'}} \bigr\rangle
	= \delta_{ll'} \, \delta_{mm'} \int \! \xi(\R) \, \conj{D^l_{ss'}(\R)} \, \D\R \;,
\end{equation}
where the symmetry $(-1)^{s+s'} \, D^l_{-s,-s'}(\R) = \conj{D^l_{ss'}(\R)}$ was used to simplify the result.
Finally, rewriting the Wigner (large) $D$-matrix in terms of the Wigner (small) $d$-matrix \citep{1967JMP.....8.2155G},
\begin{equation}
	D^l_{mm'}(\phi, \theta, \gamma)
	= \E^{\I \, m \, \phi} \, d^l_{mm'}(\theta) \, \E^{\I \, m' \, \gamma} \;,
\end{equation}
inserting the functional form~\eqref{eq:xi-rel} of the correlation function, and evaluating the trivial integrals, yields the main result of the spectral representation for SWSRF,
\begin{equation}\label{eq:sr}
	\bigl\langle X_{lm} \, \conj{X'_{l'm'}} \bigr\rangle
	= \delta_{ll'} \, \delta_{mm'} \, 2 \pi \int_{0}^{\pi} \! \xi(\theta) \, d^l_{ss'}(\theta) \sin\theta \, \D\theta \;.
\end{equation}
The differently-numbered modes of homogeneous random fields on the sphere are thereby shown to be uncorrelated for arbitrary integer spin weights.

Two related generalisations of known results follow immediately.
Comparing the result~\eqref{eq:sr} with the standard definition of the power spectrum,
\begin{equation}\label{eq:xx-cl}
	\bigl\langle X_{lm} \, \conj{X'_{l'm'}} \bigr\rangle
	\equiv \delta_{ll'} \, \delta_{mm'} \, C_l \;,
\end{equation}
yields the general expression for the modes~$C_l$ of the power spectrum for SWSRF of arbitrary integer spin weights,
\begin{equation}\label{eq:cl}
	C_l
	\equiv 2 \pi \int_{0}^{\pi} \! \xi(\theta) \, d^l_{ss'}(\theta) \sin\theta \, \D\theta \;,
\end{equation}
which agrees with the results previously found for special cases \citep{2004MNRAS.350..914C}.
Furthermore, inserting the expansion~\eqref{eq:ex-s2} into the definition~\eqref{eq:xi-s2} of the correlation function over the sphere, the expectation is moved into the sum to obtain the definition~\eqref{eq:xx-cl} of the power spectrum, while the SWSH relation~\eqref{eq:swsh-rel} can be used to absorb the exponential factors,
\begin{equation}\label{eq:xi-se}
	\xi(\beta)
	= \sum_{l,m} C_l \, {}_sY_{lm}(\phi, \theta, \alpha) \, \conj{{}_{s'}Y_{lm}(\phi', \theta', \alpha')} \;.
\end{equation}
Rewriting the second SWSH in terms of the Wigner $D$-matrix~\eqref{eq:swsh-def}, the sum over~$m$ is evaluated using the transformation law~\eqref{eq:swsh-law} in reverse,
\begin{multline}
	\sum_{m} {}_sY_{lm}(\phi, \theta, \alpha) \, \conj{{}_{s'}Y_{lm}(\phi', \theta', \alpha')} \\
	= \frac{2l + 1}{4\pi} \, D^l_{ss'}(\R(-\alpha, -\theta, -\phi) \, \R(\phi', \theta', \alpha')) \;,
\end{multline}
where the result is expressed in the Wigner $D$-matrix and simplified using its symmetries.
The product $\R(-\alpha, -\theta, -\phi) \, \R(\phi', \theta', \alpha')$ reduces to $\R(0, \beta, 0)$ by definition of the angles $\alpha, \beta, \alpha'$, so that the correlation function~\eqref{eq:xi-se} is generally recovered from the modes~\eqref{eq:cl} of the power spectrum through a simple series in~$l$,
\begin{equation}\label{eq:xi-cl}
	\xi(\beta)
	= \sum_{l} \frac{2l + 1}{4 \pi} \, C_l \, d^l_{ss'}(\beta) \;.
\end{equation}
This relation is particularly useful in applications, since it is usually the power spectrum that is predicted by theory, while the correlation function is the quantity that is more easily observed.
An efficient recursion to calculate the Wigner $d$-matrix elements for fixed $s, s'$ and increasing $l$ is given in Appendix~\ref{sec:comp}.

Finally, it can be shown that the spectral representation~\eqref{eq:sr} is not only a necessary, but also a sufficient condition for homogeneity of the fields:
Expanding the covariance~$\bigl\langle X(\R) \, \conj{X'(\R')} \bigr\rangle$ using the power spectrum~\eqref{eq:xx-cl} yields a similar expression to the correlation function~\eqref{eq:xi-se},
\begin{equation}
	\bigl\langle X(\R) \, \conj{X'(\R')} \bigr\rangle
	= \sum_{l,m} C_l \, {}_sY_{lm}(\R) \, \conj{{}_{s'}Y_{lm}(\R)} \;,
\end{equation}
with the generic rotations $\R = \R(\phi, \theta, \gamma)$ and $\R' = \R(\phi', \theta', \gamma')$ instead of the  specific angles~$\alpha$ and~$\alpha'$.
However, the expression can be rewritten using the SWSH relation~\eqref{eq:swsh-rel} to recover the correlation function~\eqref{eq:xi-se},
\begin{equation}
	\bigl\langle X(\R) \, \conj{X'(\R')} \bigr\rangle
	= \E^{-\I \, s \, (\gamma - \alpha)} \, \xi(\beta) \, \E^{\I \, s' \, (\gamma' - \alpha')} \;,
\end{equation}
which, after bringing the exponential terms to the left-hand side, is precisely the covariance~\eqref{eq:xi-s2} of a homogeneous SWSRF.

\section{Conclusion \& Discussion}
\label{sec:dis}

It was shown that the spectral representation~\eqref{eq:sr} is valid for homogeneous SWSRF with arbitrary integer spin weights, generalising the known results for the spin-$0$ and spin-$2$ cases, as well as the expression~\eqref{eq:cl} for the power spectrum, and the relation~\eqref{eq:xi-cl} to recover the correlation function.

This was achieved by extending SWSF and SWSH from the sphere to the Spin(3) group, where their transformation behaviour due to the spin weight is naturally taken into account, and showing that the modes~\eqref{eq:mo} of a SWSH expansion~\eqref{eq:ex} over Spin(3) are the same as the modes~\eqref{eq:mo-s2} of the expansion~\eqref{eq:ex-s2} over the sphere.
After relating the correlation function~\eqref{eq:xi-rel} of SWSRF over Spin(3) to the usual correlation function~$\xi(\beta)$ on the sphere, the derivation of the spectral representation is a straightforward application of the group properties of Spin(3), in particular the invariance of the Haar measure, and the SWSH transformation law~\eqref{eq:swsh-law}.

Although existing results were limited to spin-0 and spin-2 fields, there was no reason to doubt their generalisation, and the proof is provided here firstly for the sake of completeness and secondly to be able to apply the result~\eqref{eq:sr} unequivocally to any SWSRF without reference to symmetry or a decomposition in scalar-valued components.
However, perhaps more instructive than the result itself is the extension of SWSRF to Spin(3), where explicit calculations are made possible by the group structure.
As an added benefit, the approach also yields a non-geometric definition of the angles required for the practical measurement of the correlation function~\eqref{eq:xi-s2} of SWSRF, which is given in Appendix~\ref{sec:ang}.
The relation~\eqref{eq:xi-cl} offers another way to obtain the correlation function, in the form of a series in the power spectrum, and an efficient method to calculate the required terms is given in Appendix~\ref{sec:comp}.
Together, this makes it possible to measure the two-point statistics of homogeneous SWSRF of arbitrary integer spin and compare them to predictions.

\subsection*{Acknowledgements}

I would like to thank M.~Boyle, S.~Bridle, B.~Metcalf, and J.~Chluba for helpful comments and discussions.
The author acknowledges support from the European Research Council in the form of a Consolidator Grant with number 681431.


\bibliographystyle{apsrev4-1}
\bibliography{arxiv}


\appendix

\section{The Correlation Function on the Sphere}
\label{sec:ang}

For spin-weighted random fields on the sphere, the correlation function~\eqref{eq:xi-s2} is made frame-independent by rotating the fields in~$(\theta, \phi)$ and~$(\theta', \phi')$ towards the geodesic connecting the points,
\begin{equation}
	\xi(\beta)
	= \bigl\langle \E^{-\I \, s \, \alpha} \, X(\theta, \phi) \, \conj{\E^{-\I \, s' \, \alpha'} \, X'(\theta', \phi')} \bigr\rangle \;.
\end{equation}
In applications, this must be evaluated using explicit expressions for the angles~$\alpha$ and~$\alpha'$, which are not usually given.

Following Section~\ref{sec:swsrf} and Fig.~\ref{fig:rot2}, there is a direct relation between the rotations of the relative and absolute coordinate frames,
\begin{equation}\label{eq:ang}
	\R(\alpha, \beta, -\alpha')
	= \R(0, -\theta, -\phi) \, \R(\phi', \theta', 0) \;,
\end{equation}
that can serve as a purely algebraic definition of the angles~$\alpha$ and~$\alpha'$.
Multiplying out the matrices on the right-hand side, the system of equations~\eqref{eq:ang} is readily solved for $\alpha, \alpha', \beta$.
The resulting explicit forms of the angles,\footnote{%
To obtain the correct angles, the two-argument inverse tangent function should be used with the given numerator and denominator on the right-hand side.
}
\begin{gather}
	\tan\alpha
	= \frac{\sin\theta' \sin(\phi' - \phi)}{\cos\theta \sin\theta' \cos(\phi' - \phi) - \sin\theta \cos\theta'} \;, \\
	\tan\alpha'
	= \frac{\sin\theta \sin(\phi' - \phi)}{\cos\theta \sin\theta' - \sin\theta \cos\theta' \cos(\phi' - \phi)} \;,
\end{gather}
are in fact well-known expressions:
They are precisely the initial and final bearing of a vessel navigating the great circle between the points.

\section{Recovering the Correlation Function}
\label{sec:comp}

The relation~\eqref{eq:xi-cl} can be used to recover the correlation function~$\xi$ from the power spectrum~$C_l$ if a large number of Wigner $d$-matrix elements~$d^l_{mm'}(\theta)$ at increasing $l$ and fixed $m, m', \theta$ can efficiently be calculated.
This is a slightly unusual situation, since the Wigner $d$-matrix often needs to be filled out for a given angular momentum~$l$, which can be done efficiently through recursion in $m, m'$ \citep{2006JChPh.124n4115D}.
Unfortunately, that same recurrence is not particularly useful in producing the next required $d^{l+1}_{m'm}$ from the previous terms.

However, a more effective recursion is readily available.
By definition, the numbers $l, m, m'$ fulfil the inequalities $-l \le m \le l$, $-l \le m' \le l$.
There is hence a lowest value of~$l$ for which $d^l_{mm'}$ is well-defined,
\begin{equation}
	l_0 \equiv \max\{|m|, |m'|\} \;.
\end{equation}
For $l > l_0$, the Wigner $d$-matrix elements are equivalently expressed in terms of the Jacobi polynomials~$P^{(a, b)}_n$ of order $n = l - l_0$ \citep{Biedenharn:1984ex},
\begin{equation}\label{eq:jac}
    d^l_{mm'}(\beta)
    = (-1)^c \sqrt{\frac{(l-l_0)! \, (l+l_0)!}{(l-l_0+a)! \, (l-l_0+b)!}} \, \bigl(\sin\tfrac{\beta}{2}\bigr)^{a} \bigl(\cos\tfrac{\beta}{2}\bigr)^{b} \, P_{l-l_0}^{(a, b)}(\cos\beta) \;,
\end{equation}
where the non-negative integers $a, b, c$ are given by the following table of \citeauthor{Biedenharn:1984ex}:
\begin{equation}
    \begin{array}{c|ccc}
        l_0      & a         & b         & c         \\
        \hline
         m       & m  - m'   &  m  + m'  & m - m'    \\
        -m       & m' - m    & -m  - m'  & 0         \\
         m'      & m' - m    &  m  + m'  & 0         \\
        -m'      & m  - m'   & -m  - m'  & m - m'
    \end{array}
\end{equation}
There is a recurrence in~$n$ for the Jacobi polynomials \citep{1972hmfw.book.....A}, and inserting the relation~\eqref{eq:jac} yields coefficient functions $A^l_{mm'}$ and $B^l_{mm'}$,
\begin{align}\label{eq:coe}
    A^l_{mm'}
    &= \frac{(2l + 1) \bigl[l \, (l+1) \cos\theta - m m'\bigr]} {l \sqrt{(l-m+1) (l+m+1) (l-m'+1) (l+m'+1)}} \;, \\
    B^l_{mm'}
    &= \frac{(l+1) \sqrt{(l-m) (l+m) (l-m') (l+m')}} {l \sqrt{(l-m+1) (l+m+1) (l-m'+1) (l+m'+1)}} \;,
\end{align}
for a numerically stable second-order recurrence in $l$ for $d^l_{mm'}$,
\begin{equation}\label{eq:rec}
    d^{l+1}_{mm'}
    = A^{l}_{mm'} \, d^l_{mm'} - B^{l}_{mm'} \, d^{l-1}_{mm'} \;.
\end{equation}
Note that the coefficients~\eqref{eq:coe}, and hence the recurrence~\eqref{eq:rec}, do not depend on the value of $l_0$.

To find the initial values, the lowest allowed values $l = l_0$ and $l = l_0+1$ are inserted into the relation~\eqref{eq:jac}, which yields the Jacobi polynomials $P^{(a,b)}_0$ and $P^{(a,b)}_1$ with explicit expressions,
\begin{align}
    d^{l_0}_{mm'}
    &= (-1)^c \sqrt{\frac{(a + b)!}{a! \, b!}} \, \bigl(\sin\tfrac{\theta}{2}\bigr)^{a} \bigl(\cos\tfrac{\theta}{2}\bigr)^{b} \;, \\
    d^{l_0+1}_{mm'}
    &= A^{l_0}_{mm'} \, d^{l_0}_{mm'} \;.
\end{align}
Since $B^{l_0}_{mm'} = 0$ by definition, the recurrence~\eqref{eq:rec} also holds for the first term $l = l_0$, and $d^{l_0}_{mm'}$ is the only required initial value.
It can be calculated efficiently using a recursive binomial coefficient routine.

\end{document}